\documentclass[english]{article}
\usepackage[T1]{fontenc}
\usepackage[utf8]{luainputenc}
\usepackage{geometry}
\usepackage{color}
\usepackage{babel}
\usepackage{textcomp}
\usepackage{amsmath}
\usepackage{amsthm}
\usepackage{graphicx}
\usepackage{setspace}
\usepackage[unicode=true]
 {hyperref}

\makeatletter

\providecommand{\tabularnewline}{\\}

\@ifundefined{showcaptionsetup}{}{%
 \PassOptionsToPackage{caption=false}{subfig}}
\usepackage{subfig}
\makeatother

\begin{document}
\begin{doublespace}
\begin{center}
\textbf{\Large{}Seven Survival Senses: Evolutionary Training makes
Discerning Differences more Natural than Spotting Similarities}{\Large\par}
\par\end{center}

\begin{center}
\textbf{Ravi Kashyap (ravi.kashyap@stern.nyu.edu)}
\par\end{center}

\begin{center}
\textbf{City University of Hong Kong}\footnote{\begin{doublespace}
\label{The-traceable-seed}The views and opinions expressed in this
article, along with any mistakes, are mine alone and do not necessarily
reflect the official policy or position of either of my affiliations
or any other agency. The traceable seed for this paper was a talk
by Sadhguru of Isha Foundation (Vasudev 2016) in which he mentioned
that our sense organs are mainly for survival purposes. Mai Tran Anh
Tuyet, Naresh Pandey and Ablina Serikova provided excellent assistance
with conducting the experiments and other aspects of this paper. Dr.
Yong Wang, Dr. Isabel Yan, Dr. Vikas Kakkar, Dr. Fred Kwan, Dr. William
Case, Dr. Srikant Marakani, Dr. Qiang Zhang, Dr. Costel Andonie, Dr.
Jeff Hong, Dr. Guangwu Liu, Dr. Humphrey Tung and Dr. Xu Han at the
City University of Hong Kong; and numerous seminar participants provided
advice and more importantly encouragement to explore and where possible
apply cross disciplinary techniques.
\end{doublespace}
}\textbf{ }
\par\end{center}

\begin{center}
\today
\par\end{center}

\begin{center}
\par\end{center}

\begin{center}
\par\end{center}

\begin{center}
\par\end{center}

\begin{center}
\textbf{\textcolor{blue}{\href{https://doi.org/10.1080/02604027.2020.1859315}{Edited Version: Kashyap, R. (2021). Seven Survival Senses: Evolutionary Training makes Discerning Differences more Natural than Spotting Similarities.  World Futures, XX(X), XX-XX.}}}
\par\end{center}

\begin{center}
\tableofcontents{}
\par\end{center}
\end{doublespace}
\begin{doublespace}

\section{Abstract}
\end{doublespace}

\begin{doublespace}
We discuss preliminary results from two experiments and put forth
the notion that the development of sensory systems might be more geared
towards discerning differences rather than for spotting similarities.
We present the possibility that the necessity to spot differences
might have evolved to ensure the survival of the organism, which suggests
numerous other experiments to assess the response of participants
to various stimuli. We consider our present state of affairs, wherein
the need is to thrive and not merely survive, which requires us to
spot similarities around us. We provide some suggestions on how this
attribute can be developed, which includes mathematical education.
We conclude with an alternate measure for intelligence, termed the
Involvement Quotient (also, IQ), which gauges the level of involvement
of the sense organs to whatever is happening around the individual.
\end{doublespace}
\begin{doublespace}

\section{\label{sec:Senses-and-Survival}Senses and Survival}
\end{doublespace}

\begin{doublespace}
Evolutionary trajectories that helped bring about the sense organs
have been studied in great detail in various kinds of life forms (Payne
1971; Suthers 1978; Blaxter 1988; Kaas 1989; Endler 1993; Bennett
\& Cuthill 1994; Gerhardt 1994; Janssen 2004; Endler, Westcott, Madden
\& Robson 2005; Niven \& Laughlin 2008; Kaas 2008; Seehausen, ...
\& Imai 2008; Seehausen, ... \& Brelsford 2014). We pinpoint a common
theme among all these sensory developments, suggesting that they might
be more geared towards discerning differences rather than for spotting
similarities. The necessity to spot differences might have evolved
to ensure survival of the organism. For the relationship between evolution
and self-preservation or survival, see: (Doolittle \& Sapienza 1980;
Dawkins 1976; 1982; De Catanzaro 1991; Brown, ... \& Biblarz 1999.
End-note \ref{The-traceable-seed}). Any change in the environment,
coming across a different kind of food item or a different looking
creature, could be a threat indicator. When things are identical to
what we are used to, we deem that safer since the previous encounter
did not prove to be fatal. Predators (also survivors) camouflage themselves
to take advantage of this trait (Stevens \& Merilaita 2009). 

We start with a discussion of what the sense organs actually do (Section
\ref{sec:Nonsense-or-Non-sense}). We provide preliminary results
from two experiments in Section (\ref{sec:Sensing-is-More}). The
discussion suggests numerous other experiments to gauge the response
of participants (animal or human) to visual (sight), auditory (sound),
olfactory (smell), gustatory (taste), tactile (touch), vestibule (movement
and balance) and proprioceptive (body awareness or where parts of
our bodies are and how they are moving) stimuli. There is of course
much discussion about other sense organs that we possess (Craig 2003;
End-note \ref{Sense}). We venture onto other considerations regarding
the sense organs in terms of our present need wherein we are not simply
worried about mere survival (Section \ref{sec:To-Thrive-or}). Our
current necessity of having to go beyond survival requires us to spot
similarities around us and we provide pointers on how this attribute
can be developed (Sections \ref{sec:To-Thrive-or}; \ref{sec:Sleep-and-Sensing}).
We ponder about what else might happen as we continue to evolve and
the real difficulty with regards to spotting similarities (Sections
\ref{sec:Sensible-Evolution-of}; \ref{sec:Sensing-Unseen-Quirks}).
We develop an alternate measure for intelligence also called the IQ,
which is an abbreviation for the Involvement Quotient (Sections \ref{sec:A-Sensible-Measure:};
\ref{sec:Stepping-Forward-Sensibly}).
\end{doublespace}
\begin{doublespace}

\section{\label{sec:Nonsense-or-Non-sense}Nonsense or Non-sense Organs}
\end{doublespace}

\begin{doublespace}
The sense organs do an amazing job of collecting information from
the world around us. But the actual sense that happens, of all this
information that is gathered, is within the brain or the mind. An
article by Bach-y-Rita, Tyler \& Kaczmarek (2003) titled ``Seeing
with the brain'' sums it up better than any lengthy explanation we
can provide. Though right now, it seems we don't know the difference
between the mind and the brain (Churchland 1989; Oakley 2018). Hence
much of what the sense organs collect is nonsense since the brain
filters (for lack of a better word) through and picks up the relevant
bits it deems important. This clearly tells us that the sense organs
have been named incorrectly, they should be called nonsense organs.
They can also be considered non-sense organs because they are not
doing the actual sensing. They are simply collecting the stimuli,
the sensing is happening somewhere else.

An example to illustrate this is the following argument. There have
been thousands and thousands of physicists or aspiring physicists
in classrooms (and elsewhere) all around the globe, but we only have
one Albert Einstein (Fölsing 1997). The point of this argument is
that only a handful of physicists have shown us remarkable new insights
into how the physical world works when compared to the number of candidates
that are trying to glean the mysteries of the universe. All of them
were learning similar things or being trained in similar techniques.
But only some of them were able to make sense of what was happening
around them or we have been able to make sense of only what some of
them showed, did or said.

Another example is the well known story of Issac Newton and the apple.
Newton discovered gravity, which has existed since time immemorial,
only when an apple fell on his head (McKie \& De Beer 1951; Fara 1999;
End-note \ref{Newton, Gravity, Apple}). It is very likely that Newton
had seen, tasted, smelled, touched, heard an apple (or other objects)
fall before. But when it fell on his head, his brain (mind or whatever?)
finally got the message. Though the truth behind this myth can be
debated, the metaphor is relevant for us since we sometimes only notice
things that hit us smack in the face. It is unfortunate that sometimes
our sense organs fail to fully sense what is right in front of us
and we fail to appreciate (sometimes even simply acknowledge or observe?)
what is being handed to us on a platter or sent to us in a paper.

If we can categorize all (most?) human development into two broad
categories, inventions and discoveries, we could cast all progress
in terms of our sense organs. Inventions are new combinations of existing
objects to enhance (or aid) our sense organs or our ability to perceive
(and survive). Discoveries are existing phenomena that our sense organs
have newly started to perceive. The subtle point behind this grouping
of human ingenuity and its relation to our senses is that innovations
and discoveries really mean something only if we experience it. Otherwise
they are just words or symbols or stories of someone else until it
happens to us. Lane \& Flagg (2010) consider discovery, invention
and innovation as the three states of knowledge (also see, Marchetti
1980; Zahar 1983; Norman 2007). Simonton (1979) is a discussion of
the occurrence of independent contributions by two or more scientists,
which suggests that once we discover something for ourselves experientially,
we have become scientists in our own right. Once anything happens
to us, it is part of our story till then it is just his-story or history,
which is less appealing to most people than their own narrative (Kashyap
2021b). Is it any wonder that Mysteries (My-Stories) are more appealing
than Histories (His-Stories)? Montano-Rivas, McCasland, Dixon \& Bundy
(2012) is an automated approach to theorem discovery and concept innovation,
which shows that even machines have the potential to become scientists.

The real implication of this section is that, despite the seemingly
obvious fact that we are living in the same world and we might be
receiving the same (almost the same) stimuli from our surroundings,
the universe we are creating in our head might be entirely different.
This fact, (us living in the same world), becomes questionable and
open to debate based on the arguments in the rest of this article.
This could be succinctly described as we are all living in our own
worlds. Frith (2013) captures this essence in the heading of their
fifth chapter: ``Our Perception of the World is a Fantasy that Coincides
with Reality''. 

The central hypothesis of our paper considers the evolutionary reasons
regarding why our senses fail to follow certain aspects of life all
around us. We also provide suggestions on what experiments to conduct
to understand this phenomenon that ignores what is right here (perhaps
also known as ignorance), some ways to measure this tendency and ways
to overcome it. Just because we do not see a connection does not mean
that there is no connection. We need to try harder and be more open
to acknowledging knowledge everywhere or we need to accept the possibility
that something of value might have been brought to light about anything
by anyone from anywhere.
\end{doublespace}
\begin{doublespace}

\section{\label{sec:Sensing-is-More}Sensing is More Than Just Seeing}
\end{doublespace}

\begin{figure}[h]
\includegraphics[width=16cm]{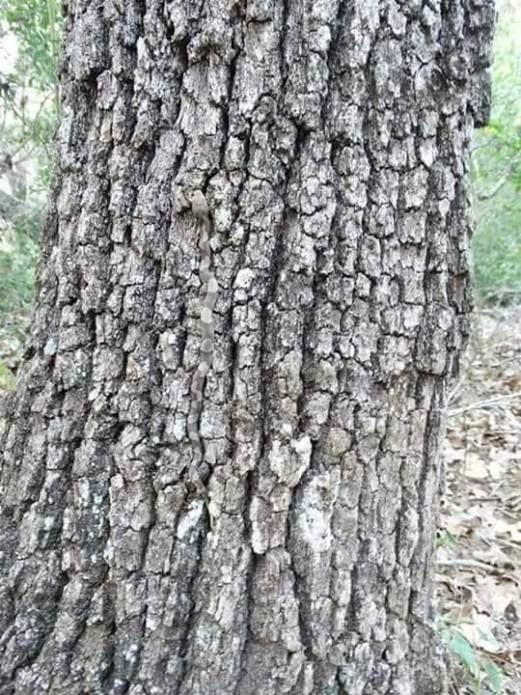}

\caption{\label{fig:See-and-Sense}See and Sense The Snake}
\end{figure}

\begin{doublespace}
We have conduced the following two experiments numerous times across
different participants. Figure (\ref{fig:Experiment-Flow-Chart})
is a flow chart of the steps involved in the two experiments. Appendix
(\ref{sec:Appendix:-Details-of}) and in particular Points (\ref{enu:The-group-sizes};
\ref{enu:Picture-One-Two}) have more details. The first test (Appendix:
Point \ref{enu:The-Snake-Experiment}) involves bringing up the picture
shown in Figure (\ref{fig:See-and-Sense}) on a screen as we give
participants instructions on various aspects of the experiment. At
this point, we have not yet pointed out that what is on the screen
is important for the experiment. Then we ask participants to describe
what they see. People see the tree, the grass and many other things.
Then we finally ask them if they saw the snake. In some cases we have
shown words on the screen on top of the figure asking ``Can you see
the snake?''. Some see it immediately after being given this clue,
many see it a few moments later, many after a much longer time. No
participant in all the tests done thus far has spotted the snake without
being told that there was a snake there. Unless someone has seen the
snake for themselves (we term this sensing the snake for lack of better
words), there is no snake for them even if we give them all the information
in the world.

\begin{figure}[h]
\includegraphics[width=15cm]{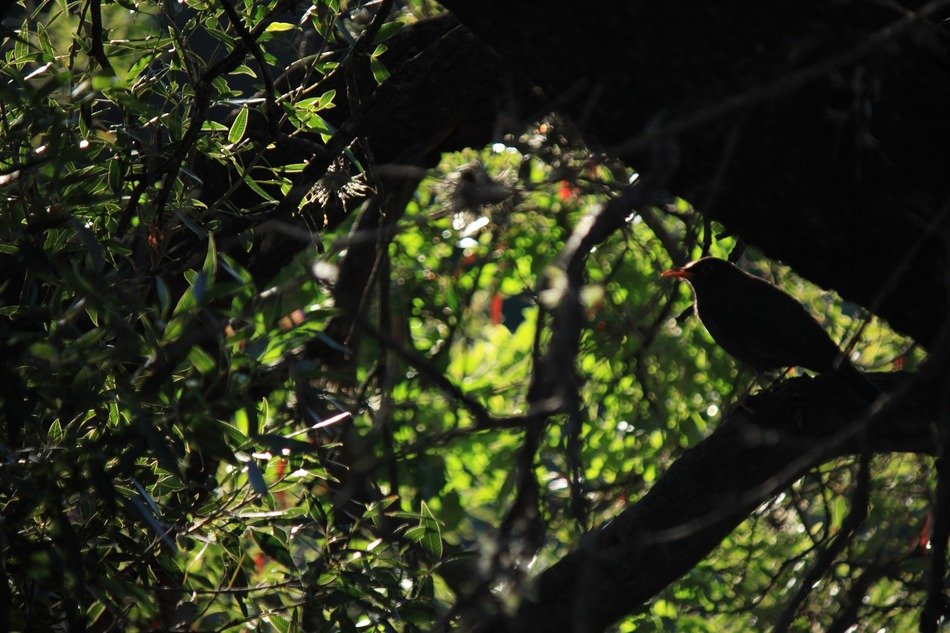}

\caption{\label{fig:Wildlife-Photographer}Wildlife Photographer}
\end{figure}

Another, similar, experiment (Appendix: Point \ref{enu:Wildlife-Photographer-Experiment})
we have conducted involves giving participants the following situation
and asking them to describe what they see in Figure (\ref{fig:Wildlife-Photographer}).
The scenario is: ``Imagine you are a wildlife photographer (Appendix:
Point \ref{enu:Wild-Life-Photographer-Definition}), walking about
in the jungles of Cambodia. Your eyes glimpse something and you take
a few photographs, one of which is shown on the screen. Why did you
take this picture?'' A small minority of the participants see the
bird in the wildlife picture within a few seconds. But the initial
reaction of many participants can be described as: peaceful, full
of nature, I took the picture because I liked the scenery etc. But
once they are prompted as to ``what is the main mission of a wildlife
photographer?'', many participants see the bird soon after.

We will publish complete results for these experiments and various
others in a separate paper. We emphasize that in the preliminary trials
conducted thus far, we did not screen the participants for any deficiencies
in eyesight including color blindness. In later rounds, we are collecting
information regarding a number of control variables such as age, sex,
education level, exam scores, nationality, etc. One other aspect to
check for and control is for the amount of sleep participants had
the night prior to this test (Appendix \ref{subsec:Participant-Control-Questionnair}).
This experiment illustrates that even though most of us receive the
same information, the extent to which we make sense of it varies and
is extremely subjective, depending on numerous factors, some innate
and fundamental, such as eyesight and some proximate and easily rectifiable,
such as lack of sleep. The results from this set of tests suggest
that we can devise numerous experiments to measure all sorts of sensory
faculties to confirm that differences are much easier to detect and
it takes practice, patience and persistence to detect similarities.
\end{doublespace}
\begin{doublespace}

\section{\label{sec:To-Thrive-or}To Thrive or Senselessly Survive}
\end{doublespace}

\begin{doublespace}
At present, the central aspect of our lives is not just to survive.
Very few of us have to worry about whether we will have food to eat,
whether we will eaten or if some conflict will end our lives. Though
we do concern ourselves with whether to eat, and many times also what
to eat and how much to eat. That we are able to spend countless hours
on such endeavors as research and writing papers shows that we, as
a civilization, are ready to thrive. We will thrive once we sense
that situations and whatever we encounter have much in common with
what we have encountered before, allowing us to utilize what we have
learnt previously. This can be succinctly descried as being able to
spot similarities.

For this to happen we have to work against the training of 4.5 billion
years, which has developed to discern differences (End-note \ref{Age-Earth}).
Here, we emphasize that evolution is nothing but training, (or survival
skills), that has kept certain genetic traits alive and passed them
onto to later generations with minor successive alterations. Shubin
(2008) suggests that evolution produced us from the first single celled
organisms that appeared 3.5 billion years ago. But perhaps we are
simply not able to sense what happened in the billion years before
that when our planet formed and until the first living organisms were
produced. We also need to think about whether the forces that created
our universe around 13.8 billion years ago, give or take a few million
years (End-note \ref{Age-Universe}), were active in creating life
as well before our planet formed and life, as we know it, took shape
upon it. Now that the question of survival is no longer an issue,
our intelligence or brain or mind, which was engaged in a very crucial
aspect of life, is suddenly devoid of purpose (or a major purpose)
bringing up numerous issues. One of which is that we imagine and create
numerous problems for ourselves, which is nothing but our intelligence
working against us due to a lack of intent or sufficient focus on
essential activities (McKnight \& Kashdan 2009).

Thriving by spotting similarities requires skills or capabilities
that the sense organs are not readily equipped to handle. This might
need adjustments, or new developments altogether, to how we perceive
the world around us. This perception can also be viewed as different
levels of awareness regarding how we receive and process the information
our sense organs collect. While evolution works at its own pace, there
might be ways in which we can develop this aptitude to spot similarity.
Being more attuned to distinguish alikeness might be a measure of
how one can move beyond survival and the tests in Section (\ref{sec:Sensing-is-More}),
might be indicative of how this feature has developed in different
individuals. 

There are age old techniques that can help with increasing our abilities
to appreciate alikeness. One such method which starts as early as
kinder-garden is known as match the following. Figure (\ref{fig:Match-The-Following})
has two examples of this. The first one (Figure \ref{fig:Somewhat-Hard-Match})
is somewhat hard and the second one (Figure \ref{fig:Relatively-Easy-Match})
is relatively easy and they are almost identical questions. A few
moments of observation will tell us that the second one is in fact
a clue to the first one. An alternative method involves merely judging
if something is true or false. Mathematical variations of our question
would involve checking if certain equations are true or false. Clearly
these could also be rephrased as match the following, which would
be a more direct test of spotting similarities. Another possibility
is the, well known, spot the difference game, wherein differences
come hidden or camouflaged among the similarities (Fukuba, etal 2009;
End-note \ref{enu:Spot-the-difference}). The similarities or differences
between spotting similarities or differences in terms of how our sensory
organs and the brain are employed need to be studied further.

If match the following sounds easy, the next suggestion is even simpler.
We simply need to be curios about everything. This is easier than
it sounds. We don't have to be curios about everything at the same
time. We just need to focus on whatever it is that is in front of
us (or around us). This also simply means that we just have to fully
engages our senses (perhaps all our senses) on whatever is around
us. If being curios sounds unnatural, perhaps we could use a few more
pointers (Section \ref{sec:Sleep-and-Sensing}).
\end{doublespace}
\begin{doublespace}

\section{\label{sec:Sleep-and-Sensing}Sleep and Sensing Symbolic Similarities}
\end{doublespace}

\begin{doublespace}
As we wait for the perfect technique(s) to thrive, it is worth meditating
upon what superior beings would do when faced with a complex situation
such as the one we are in. It is said that the Universe is but the
Brahma's (Creator's) dream (Ramamurthi 1995; Ghatage 2010). Research
(Effort / Struggle) can help us understand this world; Sleep (Ease
/ Peace of Mind) can help us create our own world. (Wilkinson 1960;
Ross 1965; Horne 1985; Tononi \& Cirelli 2006; Vassalli \& Dijk 2009;
Barnes, Schaubroeck, Huth \& Ghumman 2011) are discussions about the
functions of sleep and the adverse effects of sleep deprivation. It
is worth noting that creating this alternate universe of dreams requires
no noticeable use of the sense organs, indicating that the most powerful
senses for perception are yet to be discovered and are not oriented
to gathering signals from the external world.

\begin{figure}[h]
\includegraphics[width=17cm]{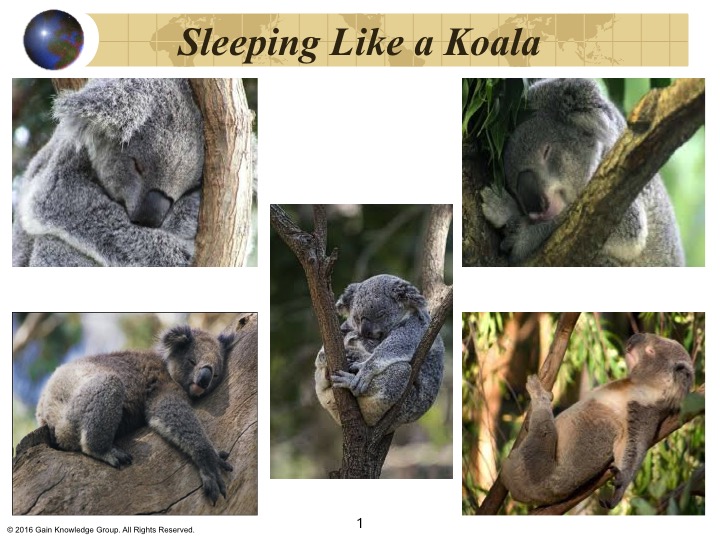}

\caption{\label{fig:Sleeping-Like-A-Koala}Sleeping Like A Koala}
\end{figure}

\begin{figure}[h]
\includegraphics[width=18cm]{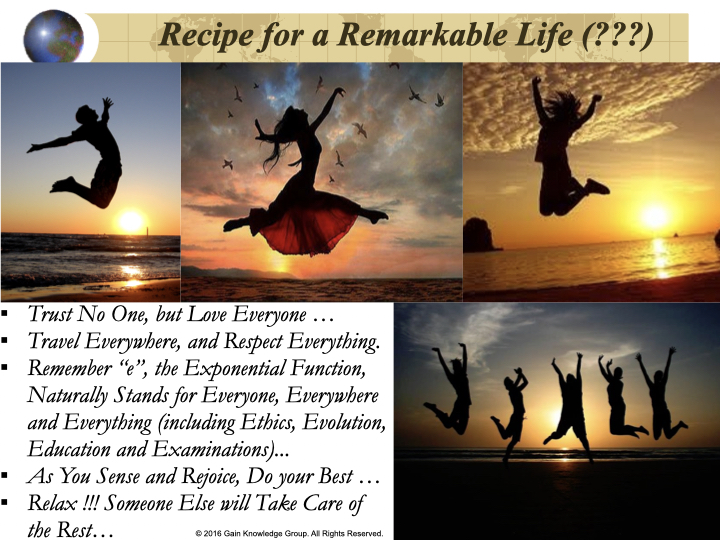}

\caption{\label{fig:Recipe-for-a}Recipe for a Remarkable Life}
\end{figure}

Native to Australia (End-note \ref{enu:Down Under}), ``Koalas spend
about 4.7 hours eating, 4 minutes traveling, 4.8 hours resting while
awake and 14.5 hours sleeping in a 24-hour period'' (Nagy and Martin1985).
The benefits of yoga on sleep quality are well documented (Khalsa
2004; Vera, ... , \& Morell 2009; End-note \ref{enu:Yoga}). A lesson
from close by and down under: ``We need to Do Some Yoga and Sleep
Like A Koala” (Figure \ref{fig:Sleeping-Like-A-Koala}). With that,
we present a list of sleeping aids in Sections (\ref{sec:Acknowledgements-and-End-notes};
\ref{sec:Sleeping-Aids}). This does not necessarily mean that we
need to sleep all the time. We need to be aware that the adequate
amount of sleep varies across organisms and individuals. Getting sleep
somewhere close to near optimal amounts could be crucial for being
involved with our environment. This also brings up the question of
what is yoga really? Yoga simply means being fully engaged, at this
present moment, in whatever we are doing. This whatever could be:
eating, drinking, laughing, smiling, talking, reading, experimenting,
swimming and so on. This is nothing but making sure that all the senses
that we control are right here and focused on whatever is around us.
Now, of course, there are always too many things around us. But we
can trust our brain or mind to zoom in on whatever we are supposed
to focus on. As they say, do your best and someone will take care
of the rest (Figure \ref{fig:Recipe-for-a}).

The sleep inducing effect of mathematics is observed in many older
humans within classrooms across the world. This therapeutic effect
is not observed in all of us, showing that there might be some of
us made for math, or, just more ready to make sense of it. Kisner,
Colby \& Borstad (2017) provide a definition of therapeutic exercises
(End-note \ref{Therapeutic_Effect}). Perhaps, babies also respond
to Mathematics by falling asleep immediately, or, in certain other
cases (not really extreme ones), by crying and displaying signs of
being frightened (possibly, getting concerned, about which world and
life they have ended up in?). More research is needed to explore this
particular connection between mathematics in the womb and beyond.
We could only find some research papers on babies and mathematics
(Dehaene 2011; Skemp 1987). The one concern we have is that if such
experiments start becoming popular, they might get classified under
torture methods due to the possibility of leading to severe childhood
trauma. This might attract the attention of human rights activists.

While the beauty and utility of mathematics are perhaps not to be
debated, the training required to discern this loveliness is non-trivial.
McAllister (2005) tries to uncover the sense of the term beauty as
it is currently used by mathematicians. (Courant \& Robbins 1941;
End-note \ref{Mathematics}) consider what mathematics really is?
It might very well be the case that the dialect of mathematics we
have adopted (with abstruse notation and omitted details, perhaps
compounded by our evolutionary training to notice apparent incongruences),
might not be that conducive to grasp meanings with. (Chinn 2013) is
a comprehensive discussion of problems when dealing with math problems
and consider how in mathematics another key communication factor is
the use of symbols. This use of symbols introduces another layer,
the need to relate symbols to vocabulary and to concepts. \textbf{\textit{This
suggests that mathematics might be about spotting similarities or
truths, despite seemingly perplex means of expression, and is ideal
training to sense the sameness in everything around us.}}

(Devlin 2000) argues that we might have an innate ability for mathematical
thinking, even though the developments in higher math have been in
the last 400 years or so, a time span which is as short as the flap
of a butterfly's wings on an evolutionary scale. (Hoffert 2009; Whiteford
2009) are discussions about teaching mathematics, supposedly the universal
language (Kashyap 2021), to students with deficiencies in the medium
of instruction such as English. This also tells us that, as we try
to consider all impediments to learning, we need to realize that the
real obstacles might simply be to check whether information is being
received in a comprehensible form (Lyon, Fletcher \& Barnes 2003;
Wong 2011).

A deeper exploration of this topic will be pursued in other venues.
But for now, it suffices for us to realize that a simple world can
be made to appear complex easily. (Kashyap 2016) consider how a few
simple deterministic rules can create a seemingly stochastic world.
So instead of puzzling over the many elements of differences that
we encounter, one of which is symbols we don't comprehend, our efforts
might be well spent marveling the creations that cross our lives and
spotting similarities among them. Let us not leave out those, so called,
non-living things. Just because we don't see (sense?) a sign of life,
does not mean that there is no life. This brings up the topic of Questions
\& Answers, Q\&A. In this case, the question is: What is a Living
creature? Biology, we suppose provides an answer, based on some Definitions
and Assumptions, D\&A (Woese, Kandler \& Wheelis 1990). But if we
change those D\&A, we might get different Q\&A, even telling us that
D\&A (and Q\&A) might be in our very DNA, which are constantly being
modified (Alberts 2017; End-note \ref{DNA}).
\end{doublespace}
\begin{doublespace}

\section{\label{sec:Sensible-Evolution-of}Evolution of Sense Organs}
\end{doublespace}

\begin{doublespace}
Most of accumulated scientific evidence points to the theory of evolution
(Darwin 1859; Futuyma \& Kirkpatrick 2009; End-note \ref{Evolution}).
If the process of evolution produced human beings from so called other
primitive life forms, this process must still be active in the descendants
of other species from which we have evolved. This is of course the
case, unless we argue and demonstrate that evolution has stopped in
some of them. If this process is still active in other life forms
it raises an important question: Are they evolving in ways we cannot
yet completely distinguish and these developments simply do not make
sense in our way of life? In addition, this process of evolution is
still at work within human beings trying to alter our sense organs
(and everything else) to the needs of today, with perhaps noticeable
(by our current measurement techniques) alternations that will appear
in the future. We discuss two instance of what might be a way forward
for the evolution of sense organs.

A simple example of a possible evolutionary adaptation is the following:
if a bomb were to explode in a crowded street, there would be chaos,
with people running all around and even on top of each other. The
following video (End-note \ref{enu:A-higher-quality}) shows a school
of fish dispersing as the film maker approaches them and as the fish
sense the shadow of the film maker. Clearly, the approaching shadow
is a sign of an impending threat. The fish react simultaneously in
a remarkable way and scatter almost immediately without running into
each other. We are continuing to run numerous experiments, as we endeavor
to find explanations and other examples in later works, to sense (not
just see?) if the fish are actually colliding and whether such a pattern
of dispersion is the norm rather than the exception. Hemelrijk \&
Hildenbrandt (2012) review model-based explanations for aspects of
the shape and internal structure of groups of fish and of birds traveling
undisturbed: that is, without predator threat. Reid, Hildenbrandt,
Padding \& Hemelrijk (2012) acknowledge that experimentally studying
the fluid dynamics of animal locomotion is difficult and time consuming,
instead they suggest the use of model based studies.

Another instance of evolutionary adaptation is based on the use of
electronic devices (phones, computers, etc.) in classrooms, while
driving and elsewhere (Rockinson-Szapkiw, Courduff, Carter \& Bennett
2013; Baker, Lusk \& Neuhauser 2012; Chang, Aeschbach, Duffy \& Czeisler
2015; Strayer, Drews \& Crouch 2006). This topic is highly debated
by policy makers and there seems to be no consensus on whether there
needs to be a limit on the use of screens. Our senses have developed
to help us survive by seeing far off things and developing the ability
to hear, even faint sounds, from all around us. But do we really need
these skills in the world today and what this world might become in
the future. Perhaps we need to see and hear better only what is right
in front of us, or at-least sense (concentrate more on) only close
by stimuli. We need to be rest assured that evolution and its endless
cycle of trials and errors will do the needful. Surely, policy needs
to be formulated when electronic device usage is harmful to someone
or their surroundings. But by forcing someone to do or not to do something
when no clear negative consequence can be sensed, we are trying to
intervene in the ways of evolution or with billions of years of wisdom
with our near sighted compassion (End-note \ref{enu:One-reason,-why}).
\end{doublespace}
\begin{doublespace}

\section{\label{sec:Sensing-Unseen-Quirks}Sensing Unseen Quirks called Quarks}
\end{doublespace}

\begin{doublespace}
We next pose the question: ``How different could everything in this
world be?'', if everything is made of the same raw material, atoms,
and just by creating different combinations of it. To create any universe
similar to the one we physically dwell in, the only building block
we need is an atom. With a certain level of understanding that overlooks
other fundamental particles: electrons, quarks and whatnots (Harari
1979; Hooft 1996; Pohl, ... \& Giesen 2010). Given this similarity
in our fundamental constitution and the marvelous differences we see
both within our behavior and around us, we need to marvel at the amazing
diversity that nature, or whoever or whatever, has created using just
different combinations of the same basic building blocks. Life would
be quite mundane without this variation. Variety is not just the spice
of life, variety is life.

All information in the vastness of the universe, can be represented
using just two symbols 0 and 1. This includes knowledge and wisdom,
without getting into the specifics of the corresponding definitions
and assumptions, or D\&A, regarding how they might differ (Kashyap
2017). Shannon (1956) showed that two symbols were sufficient to duplicate
a Turing Machine, the model for a general computer so long as enough
states were used or vice versa (Turing 1937; Woods \& Neary 2009;
End-note \ref{Turing-Machine}). The universes that we are creating,
the so called virtual worlds using computers and related technologies
are built with just two basic components, zero and one, will someday
perhaps rival our own universe in complexity (Castronova 2008; Wolf
2014). And hence these virtual worlds are a complication compared
to our physical world until we sense that zero and one might be similar
or even identical in some sense.

As we attempt to probe into the minutest depths of the world around
us and to create artificial worlds that might surpass our so called
real world, Mother Nature might take a cue and benevolently gift us
new ways of perception to succeed in those efforts. If we really want
something to happen, rest assured, evolution will develop new senses
to make it a reality, sooner or later.
\end{doublespace}
\begin{doublespace}

\section{\label{sec:A-Sensible-Measure:}A Sensible Measure: The Involvement
Quotient (IQ)}
\end{doublespace}

\begin{doublespace}
With billions and billions of years of training or preparation (a.k.a,
evolution), any individual or organism is highly likely to excel at
whatever it is supposed to do. Our current desire, almost an obsession,
to measure everything including our intelligence would be better served
if we realize that there is intelligence (even though in many cases
it cannot be easily sensed) in everyone and everything that has undergone
countless trials, errors and improvements. If every creature is capable
of intelligence perhaps what differentiates individuals in tests is
how involved they are in whatever they are doing. If someone is not
fully engaged in whatever they are doing their intelligence, which
has helped them successfully navigate numerous dangers, has not had
a chance to be utilized fully. Hence any measure of IQ (Intelligence
Quotient) is bound to be erroneous.

Also, if our body weights can change often, (and even our heights,
but more gradually, perhaps?), our IQ, must be changing even more
frequently. This is because our current measure of IQ depends on the
brain, which is changing itself much more often than our weights (Doidge
2007). Hence, we must have very different IQ levels at different times.
And if we can lose or gain weight, can we not easily do the same with
our IQ levels as well? Weinberg (1989); Bartholomew (2004) describe
the status of controversies regarding the definition of intelligence,
whether intelligence exists and, if it does, whether it can be measured,
and the relative roles of genes versus environments in the development
of individual differences in intelligence.

The above discussion suggests that we are better of first testing
how involved a person is in whatever they are doing. Checking this
would mean gauging how well the sense organs are engaged in any task
being attempted and what is the level of consciousness within the
individual to the environment in which all this is happening. The
series of tests for the sense organs, that our paper brings up, to
distinguish similarities can serve as the foundation to assess the
level of awareness towards the surroundings. We term this the Involvement
Quotient (also, IQ).
\end{doublespace}
\begin{doublespace}

\section{\label{sec:Stepping-Forward-Sensibly}Stepping Forward Sensibly}
\end{doublespace}

\begin{doublespace}
Our findings highlight that it is extremely hard to measure the actual
level of intelligence within any organism. Perhaps, with our present
perception of intelligence and prevailing tools to enhance our senses,
we can never fully do this. The next best thing we can do is gauge
the level of involvement or check how engaged the individual is with
the surroundings and how much of their sense faculties are being utilized.
Clearly, this suggests that we need to seriously rethink the current
approach to measuring intelligence (IQ: Intelligence Quotient) and
either completely substitute it or at-least complement those efforts
with assessing involvement (IQ: Involvement Quotient). The other disclosure
regarding the difficulty most of us seem to encounter while spotting
similarities suggests that a valid path ahead would involve developing
this attribute through training programs aimed at enhancing the engagement
of our senses. Such training will facilitate better perception of
alikeness. That these discoveries have numerous policy implications
for education, workforce training and for the design of a society
with greater well-being is a sensible summary of this discussion.
\end{doublespace}
\begin{doublespace}

\section{\label{sec:Acknowledgements-and-End-notes}End-notes (Some Sleeping
Aids)}
\end{doublespace}
\begin{enumerate}
\begin{doublespace}
\item \label{Sense}A sense is a physiological capacity of organisms that
provides data for perception. \href{https://en.wikipedia.org/wiki/Sense}{Senses of Living Organisms, Wikipedia Link}
\item \label{Newton, Gravity, Apple}Squirreled away in the archives of
London’s Royal Society is a manuscript containing the truth, or perhaps
the closest to one, about the apple. It is the manuscript for what
would become a biography of Newton entitled Memoirs of Sir Isaac Newton’s
Life written by William Stukeley, an archaeologist and one of Newton’s
first biographers, and published in 1752 (Stukeley 1936). Newton told
the apple story to Stukeley, who relayed it as such: “After dinner,
the weather being warm, we went into the garden and drank tea, under
the shade of some apple trees…he told me, he was just in the same
situation, as when formerly, the notion of gravitation came into his
mind. It was occasion’d by the fall of an apple, as he sat in contemplative
mood. Why should that apple always descend perpendicularly to the
ground, thought he to himself…”. \href{https://www.newscientist.com/blogs/culturelab/2010/01/newtons-apple-the-real-story.html}{Newton Gravity Apple, Link}
\item \label{Age-Earth}The age of Earth is estimated to be 4.54 \textpm{}
0.05 billion years (4.54 \texttimes{} 10\textsuperscript{9} years
\textpm{} 1\%). \href{https://en.wikipedia.org/wiki/Age_of_Earth}{Age of Earth, Wikipedia Link}
\item \label{Age-Universe}The age of the universe is the time elapsed since
the Big Bang of around 13.8 billion years. \href{https://en.wikipedia.org/wiki/Age_of_the_universe}{Age of The Universe, Wikipedia Link}
\item \label{enu:Spot-the-difference}Spot the difference is a type of puzzle
where players must find a set number of differences between two otherwise
similar images. \href{https://en.wikipedia.org/wiki/Spot_the_difference}{Spot the Difference, Wikipedia Link}
\item \label{enu:Down Under}The term Down Under is a colloquialism which
is variously construed to refer to Australia and New Zealand. \href{https://en.wikipedia.org/wiki/Down_Under}{Australia or Down Under, Wikipedia Link}
\item \label{enu:Yoga}Yoga is a group of physical, mental, and spiritual
practices which originated in ancient India. \href{https://en.wikipedia.org/wiki/Yoga}{Yoga, Wikipedia Link}
\item \label{Therapeutic_Effect}Therapeutic effect refers to the response(s)
after a treatment of any kind, the results of which are judged to
be useful or favorable. \href{https://en.wikipedia.org/wiki/Therapeutic_effect}{Therapeutic Effect, Wikipedia Link}
\item \label{Mathematics}Mathematics (from Greek máthēma, \textquotedbl knowledge,
study, learning\textquotedbl ) is the use of patterns to formulate
new conjectures. Mathematicians resolve the truth or falsity of such
by mathematical proof. When mathematical structures are good models
of real phenomena, then mathematical reasoning can provide insight
or predictions about nature. \href{https://en.wikipedia.org/wiki/Mathematics}{Mathematics, Wikipedia Link}
\item \label{DNA}Deoxyribonucleic acid (DNA) is a molecule composed of
two chains (made of nucleotides) that coil around each other to form
a double helix carrying the genetic instructions used in the growth,
development, functioning and reproduction of all known living organisms
and many viruses. \href{https://en.wikipedia.org/wiki/DNA}{DNA, Wikipedia Link}
\item \label{Evolution}Evolution is change in the heritable characteristics
of biological populations over successive generations. These characteristics
are the expressions of genes that are passed on from parent to offspring
during reproduction. \href{https://en.wikipedia.org/wiki/Evolution}{Evolution, Wikipedia Link}
\item \label{enu:A-higher-quality}A higher quality video which clearly
shows the dispersion of the fish is available upon request. The YouTube
version of this video is 10 times smaller in size than the original
video. This also shows that a lot of sense is lost in translation,
storage and transmission. \href{https://www.youtube.com/watch?v=z2ImlxLXJxM&t=0s&list=PL6FJa2nwz0MnDiUtm-WFeUw-NZR-jQqL2&index=20}{Sensible Dispersion, Video Link} 
\item \label{enu:One-reason,-why}One reason, why such unwanted outcomes
creep up is because we live in a world that requires around 2000 IQ
points to consistently make correct decisions; but the smartest of
us has only a fraction of that: (Ismail 2014; Kashyap 2021c; End-note
\ref{enu:Taleb and Kahneman discuss Trial and Error / IQ Points}).
Hence, we need to rise above the urge to ridicule the seemingly obvious
blunders of others because, without those marvelous mistakes, the
path ahead will not become clearer for us.
\item \label{enu:Taleb and Kahneman discuss Trial and Error / IQ Points}\href{https://www.youtube.com/watch?v=MMBclvY_EMA}{Nassim Taleb and Daniel Kahneman discuss Trial and Error / IQ Points, among other things, at the New York Public Library on Feb 5, 2013.}
As Taleb explains, ``... it is trial with small errors that leads
to progress ...''. That being said, if there are big errors that
might incapacitate the person trying the trial from further trials;
as long as someone else has observed the attempts with huge errors,
the rest of society benefits from it; assuming, of course, that the
big blow up has left a non-trivial portion of society intact, or at-least
not too shaken up. This concept is also an illustration of learning
from the lessons history holds for us. (Ismail 2014) mentions the
following quote from Taleb, “Knowledge gives you a little bit of an
edge, but tinkering (trial and error) is the equivalent of 1,000 IQ
points. It is tinkering that allowed the industrial revolution''.
This means that to match trial and error we need 1000 IQ points. But
trial and error could still give the wrong outcomes. So in (End-note
\ref{enu:One-reason,-why}) we make the assumption that we need 2000
IQ points to consistently make the right decisions.
\item \label{Turing-Machine}A Turing machine is a mathematical model of
computation that defines an abstract machine, which manipulates symbols
on a strip of tape according to a table of rules. \href{https://en.wikipedia.org/wiki/Turing_machine}{Turing Machine, Wikipedia Link}
\end{doublespace}
\end{enumerate}
\begin{doublespace}

\section{\label{sec:Sleeping-Aids}References (Some More Sleeping Aids)}
\end{doublespace}
\begin{enumerate}
\begin{doublespace}
\item Alberts, B. (2017). Molecular biology of the cell. Garland science.
\item Bach-y-Rita, P., Tyler, M. E., \& Kaczmarek, K. A. (2003). Seeing
with the brain. International journal of human-computer interaction,
15(2), 285-295.
\item Baker, W. M., Lusk, E. J., \& Neuhauser, K. L. (2012). On the use
of cell phones and other electronic devices in the classroom: Evidence
from a survey of faculty and students. Journal of Education for Business,
87(5), 275-289.
\item Barnes, C. M., Schaubroeck, J., Huth, M., \& Ghumman, S. (2011). Lack
of sleep and unethical conduct. Organizational Behavior and Human
Decision Processes, 115(2), 169-180.
\item Bartholomew, D. J. (2004). Measuring intelligence: Facts and fallacies.
Cambridge University Press.
\item Bennett, A. T., \& Cuthill, I. C. (1994). Ultraviolet vision in birds:
what is its function?. Vision research, 34(11), 1471-1478.
\item Blaxter, J. H. (1988). Sensory performance, behavior, and ecology
of fish. In Sensory biology of aquatic animals (pp. 203-232). Springer,
New York, NY.
\item Bommer, M., Gratto, C., Gravander, J., \& Tuttle, M. (1987). A behavioral
model of ethical and unethical decision making. Journal of business
ethics, 6(4), 265-280.
\item Brown, B. B., Clasen, D. R., \& Eicher, S. A. (1986). Perceptions
of peer pressure, peer conformity dispositions, and self-reported
behavior among adolescents. Developmental psychology, 22(4), 521.
\item Brown, R. M., Dahlen, E., Mills, C., Rick, J., \& Biblarz, A. (1999).
Evaluation of an evolutionary model of self‐preservation and self‐destruction.
Suicide and Life‐Threatening Behavior, 29(1), 58-71.
\item Castronova, E. (2008). Exodus to the virtual world: How online fun
is changing reality. Palgrave Macmillan.
\item Chang, A. M., Aeschbach, D., Duffy, J. F., \& Czeisler, C. A. (2015).
Evening use of light-emitting eReaders negatively affects sleep, circadian
timing, and next-morning alertness. Proceedings of the National Academy
of Sciences, 112(4), 1232-1237.
\item Chinn, S. (2013). The trouble with maths: A practical guide to helping
learners with numeracy difficulties. Routledge.
\item Churchland, P. S. (1989). Neurophilosophy: Toward a unified science
of the mind-brain. MIT press.
\item Courant, R., \& Robbins, H. (1941). What is mathematics?. Oxford University
Press.
\item Craig, A. D. (2003). Interoception: the sense of the physiological
condition of the body. Current opinion in neurobiology, 13(4), 500-505.
\item Darwin, C. (1859). On the Origin of Species by Means of Natural Selection
Or the Preservation of Favoured Races in the Struggle for Life. H.
Milford; Oxford University Press.
\item Dawkins, R. (1976). The selfish gene. Oxford university press.
\item Dawkins, R. (1982). The extended phenotype: The long reach of the
gene. Oxford University Press.
\item Dehaene, S. (2011). The number sense: How the mind creates mathematics.
Oxford University Press.
\item De Catanzaro, D. (1991). Evolutionary limits to self-preservation.
Ethology and Sociobiology, 12(1), 13-28.
\item Devlin, K. J. (2000). The math gene: How mathematical thinking evolved
and why numbers are like gossip. New York: Basic Books.
\item Doidge, N. (2007). The brain that changes itself: Stories of personal
triumph from the frontiers of brain science. Penguin.
\item Doolittle, W. F., \& Sapienza, C. (1980). Selfish genes, the phenotype
paradigm and genome evolution. Nature, 284(5757), 601.
\item Eisenberg, J. M. (1979). Sociologic influences on decision-making
by clinicians. Annals of Internal Medicine, 90(6), 957-964.
\item Endler, J. A. (1993). Some general comments on the evolution and design
of animal communication systems. Phil. Trans. R. Soc. Lond. B, 340(1292),
215-225.
\item Endler, J. A., Westcott, D. A., Madden, J. R., \& Robson, T. (2005).
Animal visual systems and the evolution of color patterns: sensory
processing illuminates signal evolution. Evolution, 59(8), 1795-1818.
\item Fara, P. (1999). Catch a falling apple: Isaac Newton and myths of
genius. Endeavour, 23(4), 167-170.
\item Fölsing, A. (1997). Albert Einstein: a biography. Viking. Chicago 
\item Frith, C. (2013). Making up the mind: How the brain creates our mental
world. John Wiley \& Sons.
\item Fukuba, E., Kitagaki, H., Wada, A., Uchida, K., Hara, S., Hayashi,
T., ... \& Uchida, N. (2009). Brain activation during the spot the
differences game. Magnetic Resonance in Medical Sciences, 8(1), 23-32.
\item Futuyma, D. J., \& Kirkpatrick, M. (2009). Evolution (Sinauer, Sunderland,
MA).
\item Gardner, M., \& Steinberg, L. (2005). Peer influence on risk taking,
risk preference, and risky decision making in adolescence and adulthood:
an experimental study. Developmental psychology, 41(4), 625.
\item Gerhardt, H. C. (1994). The evolution of vocalization in frogs and
toads. Annual Review of Ecology and Systematics, 25(1), 293-324.
\item Ghatage, S. (2010). Brahma's Dream. Anchor Canada, Penguin Random
House, Manhattan, New York.
\item Harari, H. (1979). A schematic model of quarks and leptons. Physics
Letters B, 86(1), 83-86.
\item Hemelrijk, C. K., \& Hildenbrandt, H. (2012). Schools of fish and
flocks of birds: their shape and internal structure by self-organization.
Interface focus, rsfs20120025.
\item Hoffert, S. B. (2009). Mathematics: The Universal Language?. Mathematics
Teacher, 103(2), 130-139.
\item Hooft, G. T. (1996). In search of the ultimate building blocks. Cambridge
University Press.
\item Horne, J. A. (1985). Sleep function, with particular reference to
sleep deprivation. Annals of clinical research, 17(5), 199-208.
\item Ismail, S. (2014). Exponential Organizations: Why new organizations
are ten times better, faster, and cheaper than yours (and what to
do about it). Diversion Books.
\item Janssen, J. (2004). Lateral line sensory ecology. In The Senses of
Fish (pp. 231-264). Springer, Dordrecht.
\item Kaas, J. H. (1989). The evolution of complex sensory systems in mammals.
Journal of Experimental Biology, 146(1), 165-176.
\item Kaas, J. H. (2008). The evolution of the complex sensory and motor
systems of the human brain. Brain research bulletin, 75(2-4), 384-390.
\item Kashyap, R. (2016). Notes on Uncertainty, Unintended Consequences
and Everything Else. Working Paper, Social Science Research Network:
https://papers.ssrn.com/sol3/papers.cfm?abstract\_id=2741040
\item Kashyap, R. (2017). Microstructure under the Microscope: Tools to
Survive and Thrive in The Age of (Too Much) Information. The Journal
of Trading, 12(2), 5-27.
\item Kashyap, R. (2021). The Universal Language: Mathematics or Music?
Journal for Multicultural Education, Forthcoming.
\item Kashyap, R. (2021b). Do Traders Become Rogues? or Do Rogues Become
Traders? The Om of Jerome and The Karma of Kerviel... Arizona State
University Corporate and Business Law Journal, 1(3), XX-XX, Forthcoming.
\item Kashyap, R. (2021c). Artificial Intelligence: A Child’s Play. Technological
Forecasting \& Social Change, Forthcoming.
\item Khalsa, S. B. S. (2004). Treatment of chronic insomnia with yoga:
A preliminary study with sleep--wake diaries. Applied psychophysiology
and biofeedback, 29(4), 269-278.
\item Kisner, C., Colby, L. A., \& Borstad, J. (2017). Therapeutic exercise:
foundations and techniques. F.A. Davis Company.
\item Lane, J. P., \& Flagg, J. L. (2010). Translating three states of knowledge--discovery,
invention, and innovation. Implementation Science, 5(1), 9.
\item Lyon, G. R., Fletcher, J. M., \& Barnes, M. C. (2003). Learning disabilities.
In E. J. Mash \& R. A. Barkley (Eds.), Child psychopathology (p. 520--586).
Guilford Press.
\item Marchetti, C. (1980). Society as a learning system: discovery, invention,
and innovation cycles revisited. Technological forecasting and social
change, 18(4), 267-282.
\item McAllister, J. W. (2005). Mathematical beauty and the evolution of
the standards of mathematical proof. In M. Emmer (Ed.), The visual
mind II (Vol. 2). Cambridge: MIT.
\item McKie, D., \& De Beer, G. R. (1951). Newton's apple. Notes and Records
of the Royal society of London, 9(1), 46-54.
\item McKnight, P. E., \& Kashdan, T. B. (2009). Purpose in life as a system
that creates and sustains health and well-being: an integrative, testable
theory. Review of General Psychology, 13(3), 242.
\item Montano-Rivas, O., McCasland, R., Dixon, L., \& Bundy, A. (2012).
Scheme-based theorem discovery and concept invention. Expert Systems
with Applications, 39(2), 1637-1646.
\item Nagy, K. A., \& Martin, R. W. (1985). Field Metabolic Rate, Water
Flux, Food Consumption and Time Budget of Koalas, Phascolarctos Cinereus
(Marsupialia: Phascolarctidae) in Victoria. Australian Journal of
Zoology, 33(5), 655-665.
\item Niven, J. E., \& Laughlin, S. B. (2008). Energy limitation as a selective
pressure on the evolution of sensory systems. Journal of Experimental
Biology, 211(11), 1792-1804.
\item Norman, A. L. (2007). Informational society: An economic theory of
discovery, invention and innovation. Springer Science \& Business
Media.
\item Oakley, D. A. (Ed.). (2018). Brain and mind. Routledge.
\item Payne, R. S. (1971). Acoustic location of prey by barn owls (Tyto
alba). Journal of Experimental Biology, 54(3), 535-573.
\item Podduwage, D. R., \& Ratnayake, P. (2020). Wildlife Photography over
Nature Photography. The International Journal of Social Sciences and
Humanities Invention, 9(9), 49-52.
\item Pohl, R., Antognini, A., Nez, F., Amaro, F. D., Biraben, F., Cardoso,
J. M., ... \& Giesen, A. (2010). The size of the proton. Nature, 466(7303),
213.
\item Ramamurthi, B. (1995). The fourth state of consciousness: The Thuriya
Avastha. Psychiatry and clinical neurosciences, 49(2), 107-110.
\item Reid, D. A., Hildenbrandt, H., Padding, J. T., \& Hemelrijk, C. K.
(2012). Fluid dynamics of moving fish in a two-dimensional multiparticle
collision dynamics model. Physical Review E, 85(2), 021901.
\item Rockinson-Szapkiw, A. J., Courduff, J., Carter, K., \& Bennett, D.
(2013). Electronic versus traditional print textbooks: A comparison
study on the influence of university students' learning. Computers
\& Education, 63, 259-266.
\item Ross, J. J. (1965). Neurological findings after prolonged sleep deprivation.
Archives of neurology, 12(4), 399-403.
\item Seehausen, O., Terai, Y., Magalhaes, I. S., Carleton, K. L., Mrosso,
H. D., Miyagi, R., ... \& Imai, H. (2008). Speciation through sensory
drive in cichlid fish. Nature, 455(7213), 620.
\item Seehausen, O., Butlin, R. K., Keller, I., Wagner, C. E., Boughman,
J. W., Hohenlohe, P. A., ... \& Brelsford, A. (2014). Genomics and
the origin of species. Nature Reviews Genetics, 15(3), 176.
\item Shannon, C. E. (1956). A universal Turing machine with two internal
states. Automata studies, 34, 157-165.
\item Shubin, N. (2008). Your inner fish: a journey into the 3.5-billion-year
history of the human body. Vintage.
\item Simonton, D. (1979). Multiple discovery and invention: Zeitgeist,
genius, or chance?. Journal of Personality and Social Psychology,
37(9), 1603-1616.
\item Skemp, R. R. (1987). The psychology of learning mathematics. Psychology
Press.
\item Stevens, M., \& Merilaita, S. (2009). Animal camouflage: current issues
and new perspectives. Philosophical Transactions of the Royal Society
of London B: Biological Sciences, 364(1516), 423-427.
\item Strayer, D. L., Drews, F. A., \& Crouch, D. J. (2006). A comparison
of the cell phone driver and the drunk driver. Human factors, 48(2),
381-391.
\item Stukeley, W. (1936). Memoirs of Sir Isaac Newton's Life. Taylor and
Francis.
\item Suthers, R. A. (1978). Sensory ecology of birds. In Sensory Ecology
(pp. 217-251). Springer, Boston, MA.
\item Tononi, G., \& Cirelli, C. (2006). Sleep function and synaptic homeostasis.
Sleep medicine reviews, 10(1), 49-62.
\item Turing, A. M. (1937). On computable numbers, with an application to
the Entscheidungsproblem. Proceedings of the London mathematical society,
2(1), 230-265.
\item Vassalli, A., \& Dijk, D. J. (2009). Sleep function: current questions
and new approaches. European Journal of Neuroscience, 29(9), 1830-1841.
\item Vasudev, J. (2016). Inner Engineering: A Yogi's Guide to Joy. Spiegel
\& Grau.
\item Venkatesh, V., Morris, M. G., \& Ackerman, P. L. (2000). A longitudinal
field investigation of gender differences in individual technology
adoption decision-making processes. Organizational behavior and human
decision processes, 83(1), 33-60.
\item Vera, F. M., Manzaneque, J. M., Maldonado, E. F., Carranque, G. A.,
Rodriguez, F. M., Blanca, M. J., \& Morell, M. (2009). Subjective
sleep quality and hormonal modulation in long-term yoga practitioners.
Biological psychology, 81(3), 164-168.
\item Weinberg, R. A. (1989). Intelligence and IQ: Landmark issues and great
debates. American Psycholo- gist, 44(2), 98.
\item Whiteford, T. (2009). Is Mathematics a Universal Language?. Teaching
Children Mathematics, 16(5), 276-283.
\item Wilkinson, R. T. (1960). The effect of lack of sleep on visual watch-keeping.
Quarterly Journal of Experimental Psychology, 12(1), 36-40.
\item Woese, C. R., Kandler, O., \& Wheelis, M. L. (1990). Towards a natural
system of organisms: proposal for the domains Archaea, Bacteria, and
Eucarya. Proceedings of the National Academy of Sciences, 87(12),
4576-4579.
\item Wolf, M. J. (2014). Building imaginary worlds: The theory and history
of subcreation. Routledge.
\item Wong, B. (Ed.). (2011). Learning about learning disabilities. Elsevier.
\item Woods, D., \& Neary, T. (2009). The complexity of small universal
Turing machines: A survey. Theoretical Computer Science, 410(4-5),
443-450.
\item Zahar, E. (1983). Logic of discovery or psychology of invention?.
The British Journal for the Philosophy of Science, 34(3), 243-261.
\end{doublespace}
\end{enumerate}
\begin{doublespace}

\section{\label{sec:Appendix:-Details-of}Appendix: Details of Experiments }
\end{doublespace}
\begin{doublespace}

\subsection{\label{subsec:Experiment-Setup-and}Experiment Setup and Summary}
\end{doublespace}

\begin{figure}[h]
\includegraphics[width=17cm]{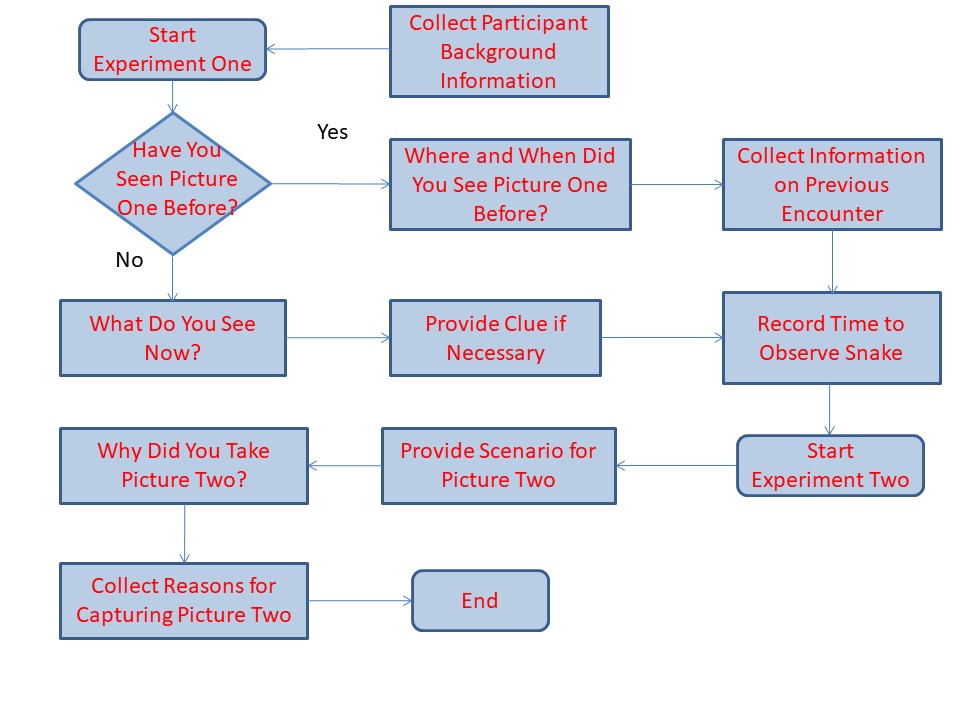}

\caption{\label{fig:Experiment-Flow-Chart}Flow Chart of Experiments}

\end{figure}

\begin{doublespace}
\renewcommand{\labelenumi}{[\Alph{enumi}]}
\end{doublespace}
\begin{enumerate}
\begin{doublespace}
\item \label{enu:The-group-sizes}The group sizes ranged from 25 to 60.
We are performing more group experiments and also trying this on individuals
after collecting information for a set of control questions (Appendix
\ref{subsec:Participant-Control-Questionnair}). For all the group
experiments, all human and animal subjects were engaged in their natural
environment, hence the only indication given that this was an experiment
was in mentioning that this could be a topic of a research paper.
Individual testing will explicitly make clear that this is an experiment
and specific protocols for human and animal testing will be followed.
If necessary, participant signature and email address or phone number
will be collected as indication of their consent to voluntarily participate
in this experiment. We also record the date, time, city and country
regarding the conduct of the individual test. The data collection
from the individual experiments will be subjected to numerous quantitative
analysis techniques. The objective of the analysis would be establish
the factors that indicate a higher degree of involvement of the individual
with the surroundings and the activities they are performing. This
will allow us to develop programs at all levels of education to ensure
individuals can perform at the peak of their abilities.
\item \label{enu:Picture-One-Two}Picture one in the flowchart in Figure
\ref{fig:Experiment-Flow-Chart} refers to Figure \ref{fig:See-and-Sense}.
Picture two in the flowchart in Figure \ref{fig:Experiment-Flow-Chart}
refers to Figure \ref{fig:Wildlife-Photographer}.
\item \label{enu:The-Snake-Experiment}The response time to spot the snake
varies from a couple of seconds to almost many minutes. When this
experiment was conducted in groups and once someone has seen the snake
and mentions it, some other members say that they see it too. If they
are then questioned, where is the snake? They point to a wrong location
in the picture. This tells us how we sometimes make decisions based
on what we see others doing or saying, which we believe is correct,
even though it has not come to us experientially or something that
has resonated with our senses, perception and understanding. This
effect can be broadly identified as having some overlaps with the
many studies conducted under the label: Peer Pressure (Eisenberg 1979;
Brown, Clasen \& Eicher 1986; Bommer, Gratto, Gravander \& Tuttle
1987; Venkatesh, Morris \& Ackerman 2000; Gardner \& Steinberg 2005).
\item \label{enu:Wildlife-Photographer-Experiment}A small minority of the
participants see the bird in the wildlife picture within 5 seconds. 
\item \label{enu:Wild-Life-Photographer-Definition}We use the definition
of Wildlife Photography found at this link: \href{http://rps.org/news/2014/may/nature-definition-agreed}{Definition of Wildlife Photopgraphy,  Link}
(also see, Podduwage \& Ratnayake 2020; \href{https://en.wikipedia.org/wiki/Wildlife_photography}{Wildlife Photography, Wikipedia Link}).
The Photographic Society of America (PSA) which represents 6500 members
and 470 camera clubs, the Fédération Internationale de l'Art Photographique
(FIAP) which represents more than 85 national associations and The
Royal Photographic Society (RPS) with over 11,000 UK and international
members will all use the same definition for nature and wildlife categories
for their respective competitions and exhibitions. The new definition
will come in to effect from 1 January 2015.
\end{doublespace}
\end{enumerate}
\begin{doublespace}
\renewcommand{\labelenumi}{[\arabic{enumi}]}
\end{doublespace}
\begin{doublespace}

\subsection{\label{subsec:Participant-Control-Questionnair}Participant Control
Questionnaire}
\end{doublespace}

\begin{doublespace}
It is optional for the participants to provide answers to these control
questions (partial list).
\end{doublespace}
\begin{enumerate}
\begin{doublespace}
\item How old are you?
\item Do you identify with any gender?
\item How many characters are there in your name?
\item What is your nationality?
\item What is your education level?
\item \label{enu:Working-Full-Time}Are you currently working full time?
\item \label{enu:Working-Part-Time}Are you currently working part time?
\item If yes to \ref{enu:Working-Full-Time} or \ref{enu:Working-Part-Time},
what is the nature of your work?
\item \label{enu:Studying-Full-Time}Are you currently studying full time?
\item \label{enu:Studying-Part-Time}Are you currently studying part time?
\item \label{enu:What-Studying}If yes to \ref{enu:Studying-Full-Time}
or \ref{enu:Studying-Part-Time}, what are you studying?
\item How many languages do you speak?
\item How many languages can you read and write?
\item What is your mother tongue?
\item Which language are you most comfortable speaking?
\item Which language are you most comfortable reading and writing?
\item Which sports do you play?
\item How many times a week?
\item How long do you play each time in hours?
\item Do you go to the gym?
\item How many times a week?
\item How long do you go to the gym each time in hours?
\item How many countries have you traveled to?
\item How many hours of sleep do you get on average?
\item How many hours of sleep did you get yesterday?
\item Please provide any Exam Score {[}International English Language Testing
System (IELTS), Graduate Management Admission Test (GMAT), Test of
English for International Communication (TOEIC), Test of English as
a Foreign Language (TOEFL), Other{]}.
\end{doublespace}
\begin{enumerate}
\begin{doublespace}
\item Exam and Score are noted down.
\end{doublespace}
\end{enumerate}
\begin{doublespace}
\item \label{enu:Have-you-taken}Have you taken any course in mathematics
or the latest math course you took?
\item If yes to \ref{enu:Have-you-taken}, Was this during high school,
undergraduate or graduate level?
\item If yes to \ref{enu:Have-you-taken}, What was your grade in that mathematics
course?
\item If yes to \ref{enu:Have-you-taken}, What was the maximum possible
grade for that math course?
\item If yes to \ref{enu:Have-you-taken}, What was your overall GPA (Grade
Point Average) for the program the above course was part of? If not
completed, latest GPA.
\item If yes to \ref{enu:Have-you-taken}, What is the maximum GPA possible
for the program the above course was part of?
\item What was your major at the last institution or school you attended?
Or, What is your major at the present institution or school you are
attending?\footnote{Despite the similarity between this question and question \ref{enu:What-Studying},
we wish to highlight that the answers could be different? There could
be a difference between what we are studying and our majors! Also,
for someone currently studying, the answer to this question could
be the institution attended before attending this present one.}
\item What was the name of the last university or school you attended? Or,
What is the name of the present university or school you are attending?
\item How many hours do you spend on your phone everyday?
\item How many hours do you spend on your computer everyday?
\item How many hours of television do you watch everyday?
\item How many hours do you read a regular printed book everyday?
\item Do you have color blindness?
\item \label{enu:Glasses-Contact-Lens}Do you wear contact lens or glasses?
\item If yes to \ref{enu:Glasses-Contact-Lens}, What is the reason for
wearing glasses or contact lens?
\item Have you had corrective eye surgery?
\item Do you have any other eye sight defects?
\item Have you ever been operated under the use of general anesthesia?
\item Do you like mathematics?
\item What time of the day do you feel most energetic usually? (Morning,
Afternoon, Evening or Night)
\end{doublespace}
\end{enumerate}
\begin{doublespace}

\section{Figures}
\end{doublespace}

\begin{doublespace}

\begin{figure}[h]
\subfloat[\label{fig:Somewhat-Hard-Match}Somewhat Hard]{{\LARGE{}}%
\begin{tabular}{|c|c|}
\hline 
{\LARGE{}$\log$12} & {\LARGE{}$\theta\alpha@\beta$}\tabularnewline
\hline 
\hline 
{\LARGE{}$\log$200} & {\LARGE{}$\alpha@\theta$}\tabularnewline
\hline 
{\LARGE{}$\log$$\mathbf{\frac{14}{3}}$} & {\LARGE{}$\alpha\lambda\beta@\delta$}\tabularnewline
\hline 
{\LARGE{}$\log$0.3} & {\LARGE{}$\beta\lambda\pi$}\tabularnewline
\hline 
{\LARGE{}$\log$1.5} & {\LARGE{}$\beta\lambda\alpha$}\tabularnewline
\hline 
\end{tabular}{\LARGE\par}

}\hfill{}\subfloat[\label{fig:Relatively-Easy-Match}Relatively Easy]{{\LARGE{}}%
\begin{tabular}{|c|c|}
\hline 
{\LARGE{}$\log$12} & {\LARGE{}$2x+y$}\tabularnewline
\hline 
\hline 
{\LARGE{}$\log$200} & {\LARGE{}$x+2$}\tabularnewline
\hline 
\textbf{\LARGE{}$\log$}{\LARGE{}$\mathbf{\frac{14}{3}}$} & {\LARGE{}$x-y+z$}\tabularnewline
\hline 
{\LARGE{}$\log$0.3} & {\LARGE{}$y-1$}\tabularnewline
\hline 
{\LARGE{}$\log$1.5} & {\LARGE{}$y-x$}\tabularnewline
\hline 
\end{tabular}{\LARGE\par}

}\caption{\label{fig:Match-The-Following}Match The Following: Two Examples}
\end{figure}

\end{doublespace}

\end{document}